# Learning study similarity to investigate heterogeneity in meta-analysis using LLMs and triplet loss

Kanella Panagiotopoulou[1], Harald Binder[1], Theodoros Evrenoglou[1]

[1]Institute of Medical Biometry and Statistics, Faculty of Medicine and Medical Center-
University of Freiburg, Freiburg im Breisgau, Germany

**Abstract:**

Meta-analyses of observational studies are frequently characterized by substantial between-study heterogeneity, which limits the interpretability of pooled estimates. Although conventional meta-regression can, in principle, be used to explore heterogeneity, it is often underpowered to handle multiple effect modifiers. We propose a novel framework that integrates large language models (LLMs) with deep metric learning to infer study-level similarity and uncover latent structural patterns among studies prior to meta-analysis. Study-level clinical and methodological characteristics were processed by an LLM to generate relative similarity constraints in the form of study triplets (anchor, similar, dissimilar). These triplets were constructed by treating each study as an anchor and comparing it with pairs of other studies to identify, in each instance, the study most similar to the anchor. Then, the triplets were used into an embedding model trained with triplet loss; a deep learning approach widely applied in fields such as computer vision to learn similarities between data points. In our context, this algorithm learns an embedding space where clinically and methodologically similar studies are clustered together. We apply our framework to a published meta-analysis dataset of 58 observational studies comparing cognitive outcomes between preterm- and term-born children. Subsequently, we fit meta-analysis models within the identified study clusters and compare the results with those of the overall analysis. Results suggested three clusters two of which retained considerable between-study heterogeneity. The remaining cluster comprised the most homogeneous group of studies and exhibited a more extreme pooled effect estimate together with a narrower prediction interval compared with the overall analysis. This work presents a novel approach for exploring heterogeneity in meta-analysis by incorporating available covariate information prior to model fitting. By transforming unstructured study characteristics into a similarity space, our framework enables the identification of coherent subgroups. This approach extends beyond current practice, where heterogeneity is typically investigated only after the primary analysis. Overall, the proposed framework supports more precise inference, thereby enhancing decision-making in the presence of heterogeneous real-world evidence.

## 1. Introduction

Between-study heterogeneity is a major challenge in meta-analysis with important implications for evidence synthesis and decision making[1–4]. When effect estimates vary substantially across studies, the interpretation of a single pooled effect becomes problematic, as the summary estimate may fail to represent any clinically meaningful population or setting, limiting its practical relevance[5–7]. In such contexts, making inferences solely based on the pooled estimate and its associated confidence interval can be misleading or insufficient, as important differences across studies are obscured. In addition, although prediction intervals aim to capture the range of plausible effects in future studies[8], they often become too wide, to provide actionable guidance. Consequently, substantial heterogeneity reduces the precision, reliability and clinical interpretability of meta-analytic conclusions.

These challenges become even more pronounced when meta-analyses incorporate non-randomized evidence, such as observational studies (OSs), which are increasingly used to inform real-world medical decisions[9,10]. While observational evidence enhances generalizability by reflecting routine practice and more diverse populations, it also introduces multiple sources of heterogeneity, including differences in study design, population characteristics, exposure definitions, and analytical approaches. The absence of randomization further increases susceptibility to confounding and bias, further complicating interpretation. As a result, meta-analytic findings from observational studies are frequently highly variable[11], making it difficult to determine whether differences in results reflect true variation in effects across contexts or underlying biases[12]. This uncertainty poses a key challenge for practitioners and policy stakeholders, as it complicates the assessment of whether, and to what extent, the available evidence is generalizable across healthcare settings and populations.

Although random-effects models incorporate heterogeneity through the between-study variance component, they rely on the assumption that study-specific effects are exchangeable draws from a common (typically normal) distribution[13]. This assumption may be overly restrictive for non-randomized evidence, where heterogeneity often reflects complex clinical and methodological differences across studies[14]. Alternatively, while more flexible models that relax the random-effects normality assumption provide additional insights into the underlying structure of the data, most of them remain largely descriptive and do not explicitly account for different sources of variability across studies[5]. Moreover, meta-regression and subgroup analyses aim to explain heterogeneity by incorporating study-level covariates, but their usefulness depends on the availability and quality of relevant effect modifiers, which are frequently missing or inconsistently reported[13,15]. In practice, these approaches are often underpowered when multiple covariates are considered simultaneously[16]. Consequently, despite methodological advances, heterogeneity may remain unexplained, making it more difficult to derive practical insights that are relevant for medical decision-making.

These limitations motivate the need for data-driven approaches capable of elucidating complex relationships among study-level characteristics. In particular, in observational studies, heterogeneity often arises from the joint contribution of multiple clinical and methodological features, whose combined structure is difficult to account for using conventional meta-analytic frameworks. Recent developments in machine learning and representation learning provide new opportunities to model such complex study-level relationships in a more flexible and data-adaptive manner[17]. We introduce a novel framework that integrates large language models (LLMs) with deep metric learning[18] to explore heterogeneity in meta-analyses of observational studies. Specifically, an LLM is used to extract and encode rich study-level information, which is then leveraged within a representation learning method to capture similarities between studies. This approach enables the identification of more homogeneous groups of studies prior to meta-analysis, which may support more reliable and interpretable meta-analytic findings. The remainder of the paper is structured as follows: Section 2 describes the proposed methodological framework, Section 3 presents an application to a real-world case study, and Section 4 discusses implications, limitations, and future directions.

## 2. Methods

### 2.1 Study representation framework

To learn similarity relationships between studies based on their clinical and methodological characteristics, we developed a two-stage framework consisting of (i) triplet generation using a large language model and (ii) representation learning via a soft ordinal embedding model, where the generated triplets inform training through a triplet loss function. This approach yields a $d$-dimensional vector representation (embedding)[17,19]of each study, where distances between vectors reflect relative similarity: studies with similar characteristics are positioned closer together, while more dissimilar studies are placed farther apart. The resulting embedding space provides a data-driven representation of study relationships that can be used to identify subgroups and latent structure prior to meta-analysis.

#### 2.1.1 Triplet generation using LLM

Triplets of studies were constructed to encode relative similarity relationships. Each triplet takes the form $(a, p, n)$, where $a$ is the anchor study, $p$ is the positive study considered more similar to the anchor, and $n$ is the negative study considered less similar to the anchor[18,20]. To determine the relative study similarity within each triplet, we employed an LLM. For each anchor study, two candidate studies were sampled from the remaining dataset, and the model was prompted to identify which candidate was more similar to the anchor according to the available clinical and methodological characteristics. To assess whether the resulting triplets were meaningful, the LLM was also asked to provide a brief explanation for each triplet selection. These explanations were used for qualitative evaluation, examining whether the model's reasoning underlying the triplet selection was reasonable and coherent. We first generated a large sample of triplets and then randomly selected a subset following the recommendations of Jamieson and Nowak[21]. Specifically, the number of retained triplets was determined according to:

$$\lambda m d log(m) \tag{1}$$

where $m$ represents the number of studies, $d$ the embedding dimension and $\lambda \epsilon \mathbb{N}$ is a scaling factor referred to as the triplet multiplier. Equation (1) corresponds to a known lower bound on the number of informative triplets required to learn a stable embedding representation. A fixed seed was also used to ensure reproducibility.

#### 2.1.2 Representation learning via soft ordinal embedding model trained with triplet loss

To learn a representation guided by the LLM-generated triplets, we used the soft ordinal embedding (SOE) model[22]. In a systematic evaluation of ordinal embedding algorithms, SOE demonstrated the most stable and consistently better performance across a wide range of scenarios[23]. The objective of this approach is to map each study to a point in a $d$-dimensional Euclidean space such that the relative similarity constraints encoded by the triplets are satisfied as closely as possible. Let $X \in \mathbb{R}^{m \times d}$ denote the output matrix of study embeddings, where the $i^{th}$ row, $x_i \in \mathbb{R}^d$, represents the embedding vector of study $i$, for $i = 1, \ldots, m$ studies. For each triplet of studies $(a, p, n)$, the model encourages study $a$(anchor) to be closer to study $p$(positive) than to study $n$(negative) in the embedding space. This is achieved by minimizing the triplet loss function[24]:

$$\mathcal{L}_{a,p,n} = \max\left(0, \delta + \left\|x_a - x_p\right\|_2 - \|x_a - x_n\|_2\right)$$

where $x_a, x_p, x_n \in \mathbb{R}^d$ denote the embeddings of studies $a, p$ and $n$, respectively, $\|\cdot\|_2$ is the Euclidean norm, and $\delta > 0$ is the margin parameter. This loss encourages that the anchor-positive distance is smaller than the anchor-negative distance by at least a margin $\delta$. Only triplets that violate this constraint contribute a positive loss, which promotes separation between similar and dissimilar studies in the embedding space. The overall loss function is obtained by summing $\mathcal{L}_{a,p,n}$ over all triplets. The embedding matrix $X$ is randomly initialized and estimated via stochastic gradient optimization, using mini-batches of triplets. At each iteration, gradients are computed only with respect to the embeddings involved in the current batch, enabling efficient scaling to large numbers of triplets. Optimization was performed for a fixed number of epochs, and the embedding configuration associated with the lowest proportion of violated triplets (triplet error) was retained as the final solution. Through this optimization process, the learned study embeddings encode the relative similarity relationships specified by the LLM-generated triplets, with more similar studies positioned closer together in the embedding space.

To monitor training progress, we tracked both the average triplet loss and the triplet error over time. The triplet loss measures the average magnitude by which the margin constraint is violated, while the triplet error measures the proportion of triplets that violate the constraint, irrespective of the magnitude of the violation. Together, these metrics provide interpretable measures of how well the learned embeddings satisfy the similarity constraints.

### 2.2 Subgroup identifications and meta-analysis

The learned study embeddings were clustered using the $k$-means algorithm, one of the most widely used partitioning methods for grouping observations into $k$ pre-specified, mutually exclusive groups (i.e. $k$ clusters)[25]. Let $\mu_j$ represent the center of cluster $C_j$, for $j = 1, \dots, k$, where $x_i \in \mathbb{R}^d$ denotes the embedding vector of study $i$. The $k$-means algorithm partitions the embeddings into $k$ clusters by minimizing the total within-cluster variation, defined as:

$$\sum_{j=1}^{k} \sum_{x_i \in C_j} d(x_i, \mu_j) \tag{2}$$

where $d(\cdot,\cdot)$ denotes a distance measure between observations and cluster centers. Minimizing Equation (2) encourages observations within the same cluster to be as similar as possible, thereby promoting separation between clusters. The main challenge in applying $k$-means clustering is selecting the appropriate number of clusters $k$. In the absence of prior knowledge regarding the underlying cluster structure of the data, the optimal value of $k$ can be approximated using the elbow method. Specifically, the within-cluster variation is calculated across a range of candidate values for $k$ and visualized using an elbow plot. Although the within-cluster variation generally decreases as $k$ increases, the rate of decrease eventually slows. The value of $k$ corresponding to this inflection point, commonly referred to as the "elbow," was considered indicative of an appropriate number of clusters.

Subsequently, subgroup analysis based on the identified clusters was conducted to investigate whether the resulting clusters could provide meaningful insights regarding the between-study heterogeneity.

### 2.3 Sensitivity analyses

In order to assess whether the learned embeddings and the overall conclusions were sensitive to the specific set of triplets used, we constructed alternative triplet datasets of the same size as in the main analysis by drawing different subsets from the original large triplet set using different seeds. This procedure varied the triplet composition while keeping the total number of triplets fixed. In addition, we examined different $k$ values to assess the robustness of the clustering results to the chosen number of clusters. Finally, we investigated the sensitivity of the results to the hyperparameters in Equation (1), which determines the number of retained triplets. Specifically, we varied the triplet multiplier $\lambda$ to examine the impact of using fewer or more triplets on the stability of the learned embeddings. We also increased the embedding dimension $d$ to assess whether higher-dimensional embeddings yield more informative representations of study similarity or instead introduce noise and reduce interpretability.

## 3. Real world case study

### 3.1 Motivating meta-analysis dataset

We applied the proposed framework to a published meta-analysis of 58 cohort studies comparing the Intelligence Quotient (IQ) differences between children born extremely (EPT) or very preterm (VPT) and those born at full term. IQ was assessed between 4 and 7 years of age. EPT was defined as birth before 28 weeks of gestation and/or birth weight below 1000g, whereas VPT was defined as birth before 32 weeks of gestation and/or birth weight below 1500g[26]. Three studies from the original dataset of 61 studies were excluded as they reported multiple estimates from gestational age subgroups[27–29]. The outcome measure was standardized mean difference (SMD). The estimated between-study heterogeneity was $\tau^2 = 0.11$, which remained largely unexplained after accounting for the most important study-level covariates using traditional meta-regression ($\tau^2 = 0.08$). Supplementary Table 1 summarizes the study-level characteristics included in the analysis. Previous empirical investigations of heterogeneity in meta-analyses with continuous outcomes measured using SMDs have reported substantially smaller typical values, with median $\tau^2$ estimate of 0.03[30]. Therefore, the observed heterogeneity in the present meta-analysis may be considered moderate to high relative to what is commonly encountered in practice.

### 3.2 Implementation

The triplet construction procedure described in Section 2 was implemented using a custom deployment of an LLM model from the GPT-5.4 family (openai/gpt-5.4-llmlb) developed by OpenAI[31]. For each anchor study, two candidate studies were randomly sampled from the remaining studies, and the model was asked to identify which candidate was more similar to the anchor according to predefined clinical and methodological characteristics. The relevant prompt template is provided in Supplementary Appendix 1.

Equation (1) yielded a total of 1160 randomly selected triplets, corresponding to triplet multiplier of $\lambda = 2$ and embedding dimension of $d = 2$. The choice of $\lambda$ was guided by Vankadara et al[23], while the embedding dimension was set to $d = 2$ given the relatively small number of studies ($m = 58$ studies) and to facilitate visualization and interpretation of the learned representations. The triplet loss function was implemented with a margin parameter $\delta = 1$ to encourage separation between similar and dissimilar studies. Optimization was performed using the Adam optimizer[32] with a learning rate of 0.01, over 300 epochs, and with a mini-batch size of 128 triplets. To maintain consistency with the SOE model, $k$-means clustering was applied using standard squared Euclidean distance, such that in Equation (2), $d(\cdot,\cdot) = \|x_i - \mu_j\|_2^2$. The optimal number of clusters $k$, was determined using the elbow method which indicated an inflection point at $k \approx 3$ (Supplementary Figure 1). Consequently, this value was used in the primary analysis, while $k = 2,4$ and 5 were examined in sensitivity analyses. We also assessed the robustness of the results using three different seeds (20, 50, and 100), each generating a distinct subset of triplets from the original triplet set. Additional sensitivity analyses were performed across alternative tuning parameters settings, with $\lambda \in \{1,2,4\}$ and $d \in \{2,5,10\}$.

All methods were implemented in Python (version 3.12.0), whereas subgroup and meta-regression analyse s were conducted in R (version 4.5.3).

## 3.3 Results

### 3.3.1 Main findings

The LLM-generated triplets and their corresponding similarity explanations can be found in Supplementary Appendix 2. Manual review of a subset of the generated LLM explanations indicated that the inferred similarity relationships were reasonable and coherent. Representative examples are also provided in the Supplementary Table 2.

Overall, the SOE model demonstrated stable training behavior. The proportion of violated triplets decreased rapidly during the initial training epochs and stabilized at a low level, indicating that the learned embeddings increasingly satisfied the ordinal constraints derived from the LLM-generated triplets (Figure 1a). The minimum triplet error was below 0.10 and was achieved at approximately 150 epochs, after which it remained relatively stable with only a slight increase. The triplet loss decreased monotonically during training and approached a stable minimum (Figure 1b). Although the triplet loss continued to decrease beyond 150 epochs, these additional reductions were relatively low and were not accompanied by further reductions in the triplet error. This suggests that subsequent optimization primarily refined the relative distances among already well-structured embeddings rather than substantially improving overall triplet satisfaction.

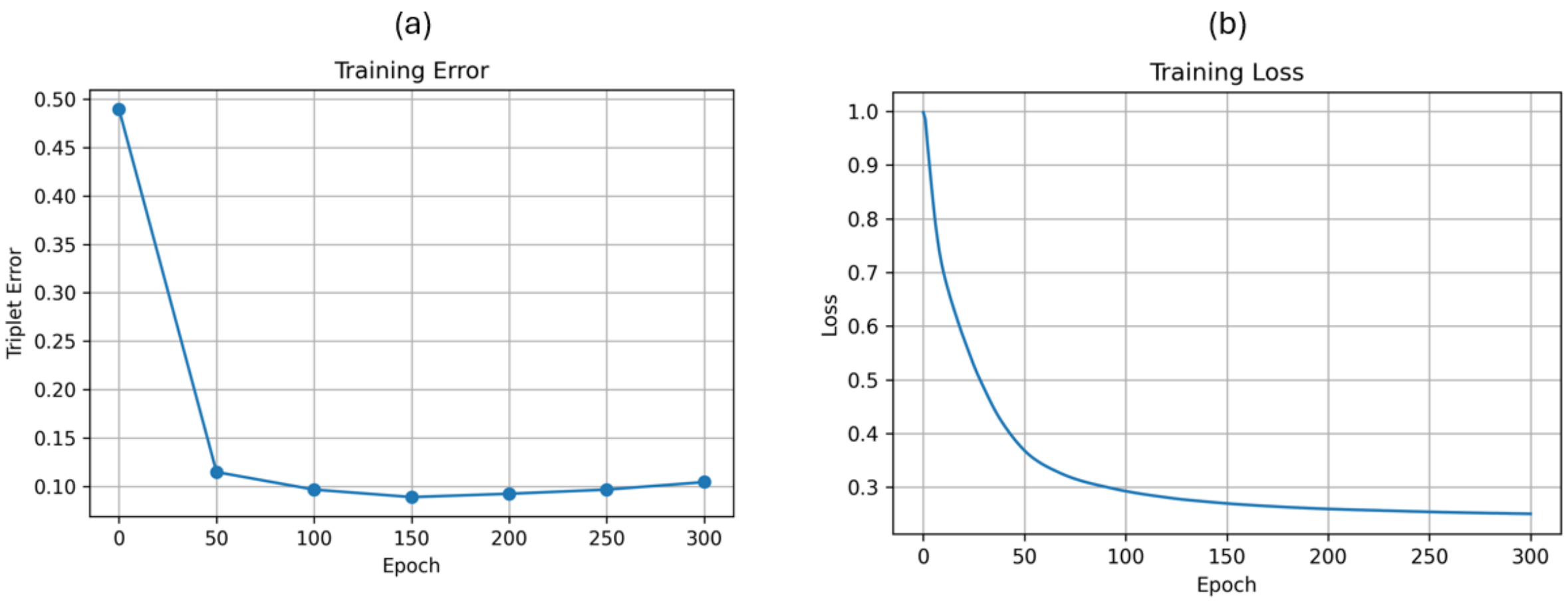

**Figure 1.** (a) Proportion of violated triplets (triplet error) and (b) triplet loss function over 300 training epochs.

Figure 2a demonstrates the two-dimensional representation of the learned embeddings for the 58 studies included in the dataset. The embedding space exhibited a non-uniform structure, with several studies forming relatively compact regions, whereas a smaller number of studies appeared more dispersed. Applying $k$-means clustering with $k = 3$ identified three reasonably distinct clusters, each containing a sufficient number of studies to support meaningful statistical inference. (Figure 2b). Subsequent subgroup analysis based on these clusters revealed differences in heterogeneity patterns and pooled effect estimates (Figure 3). In particular, Cluster 1 exhibited substantially lower between-study heterogeneity relative to both the overall analysis and the remaining clusters, with an estimated between-study variance of $\tau^2 = 0.06$ compared with $\tau^2 = 0.11$ in the overall meta-analysis. In contrast, Clusters 2 and 3 retained higher values of heterogeneity, $\tau^2 = 0.15$ and $\tau^2 = 0.11$, respectively. Cluster 1 also produced the narrowest prediction interval, indicating greater consistency in the expected range of effects for future studies within this subgroup. The pooled SMDs for Clusters 2 and 3 were broadly similar to those from the overall analysis, with substantially overlapping confidence and prediction intervals. On the contrary, Cluster 1 yielded a more extreme pooled effect estimate relative to both the remaining clusters and the analysis including all studies. Overall, these findings suggest that Cluster 1 represents a comparatively cohesive subgroup of studies characterized by greater similarity in the embedding space, lower between-study heterogeneity, and consistently more extreme effect sizes. This alignment between embedding proximity and reduced between-study heterogeneity suggests that the learned embeddings captured clinically and methodologically meaningful structure within the study-level data, enabling the identification of a more homogeneous subgroup of studies that was obscured in the overall analysis, where considerable heterogeneity was present.

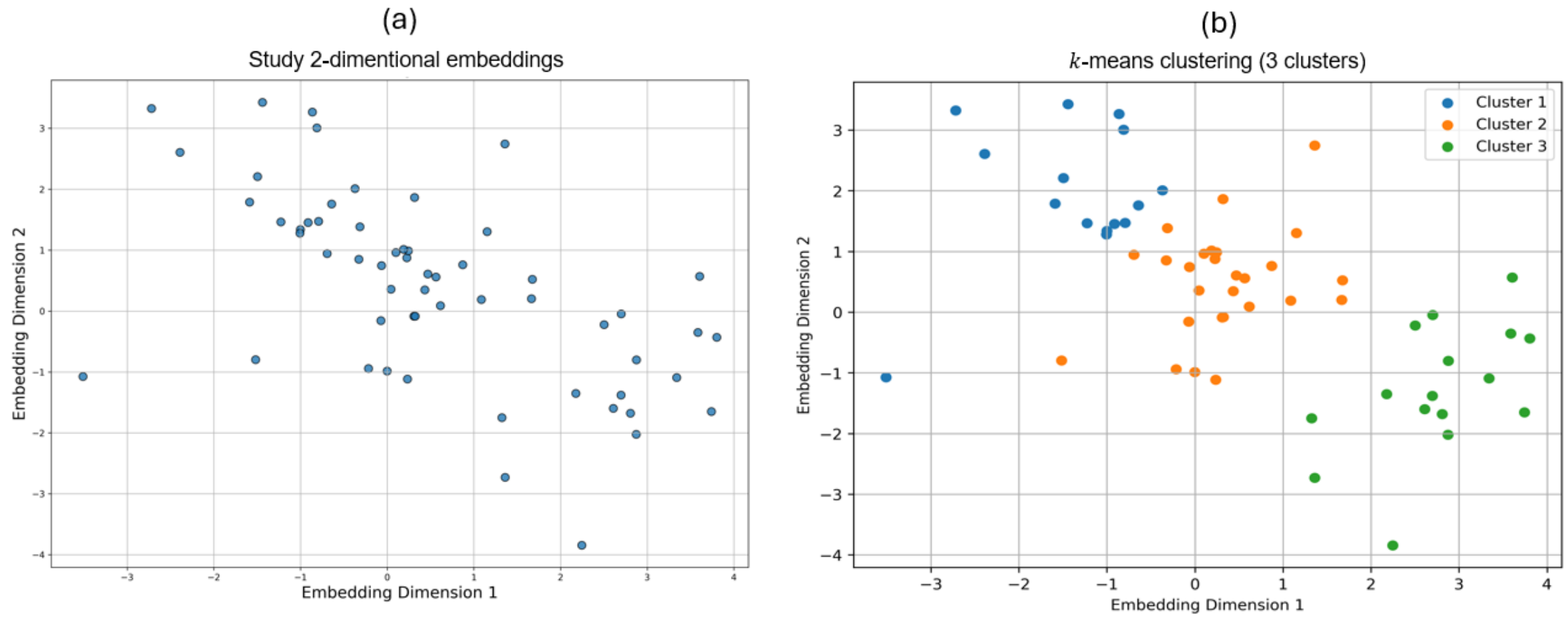

**Figure 2.** (a) Two-dimensional representation of the learned embeddings for the 58 studies included in the dataset. (b) The same two-dimensional embeddings colored according to the cluster assignments obtained using $k$-means clustering ($k = 3$).

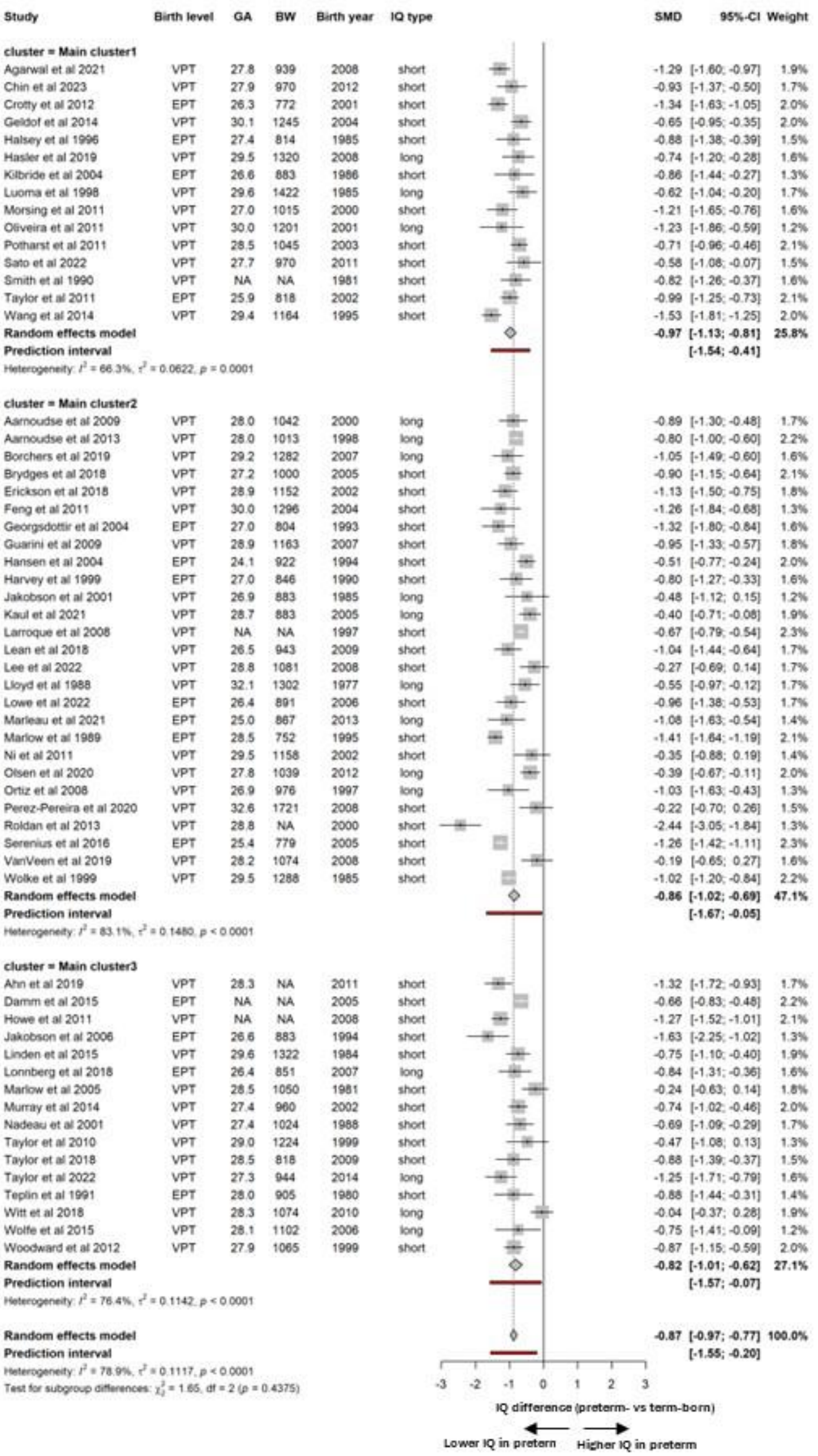

**Figure 3.** Forest plot of subgroup analyses based on clusters identified using $k$-means clustering ($k = 3$), with between-study heterogeneity ($\tau^2$) estimated using REML. BW: Birth Weight, EPT: Extremely preterm-born, IQ: Intelligence Quotient, SES: socioeconomic factors, SMDs: Standardized mean differences, VPT: Very preterm-born.

### 3.3.2 Sensitivity analysis results

Sensitivity analyses based on alternative subsets of triplets selected under different seeds produced embedding configurations that differed moderately in their spatial arrangement from the primary analysis, while still preserving the same overall geometric structure and relationships among studies (Supplementary Figure 2). Moreover, $k$-means clustering with $k = 3$ resulted in highly similar study allocations across seeds and remained largely consistent with the clustering obtained in the primary analysis, indicating that the identified subgroup structure was robust to the triplet selection process (Supplementary Figure 2 and Supplementary Table 3). Additional sensitivity analyses examining different numbers of clusters in the primary embeddings further supported the stability of the findings (Figure 4). When $k = 2$, studies were partitioned into two broad groups (Figure 4a). Increasing the number of clusters to $k = 4$ and $k = 5$ produced finer subdivisions of the embedding space, but one cluster remained essentially unchanged across these groupings and corresponded to Cluster 1 identified in the primary $k = 3$ analysis (Figure 4b–d). The consistent identification of this subgroup across both alternative embedding configurations and different clustering specifications supports the presence of a stable and distinct subgroup of comparatively homogeneous studies within the broader heterogeneous dataset. In addition, lower triplet multipliers ($\lambda = 1$) resulted in more dispersed embedding representations across dimensions, whereas higher values ($\lambda = 2$ and $\lambda = 4$) produced progressively more compact and structured study groupings, with studies concentrating around denser regions of the latent space (Supplementary Figure 3). Varying the embedding dimension from $d = 2$ to $d = 5$ led to moderate changes in the spatial arrangement and dispersion of the study embeddings across $\lambda$ values, although no fundamentally different grouping behavior emerged. In contrast, embeddings generated with $d = 10$ showed a clearer separation of studies into approximately three dense regions, especially for higher $\lambda$ values, while preserving the general structure observed in lower dimensions.

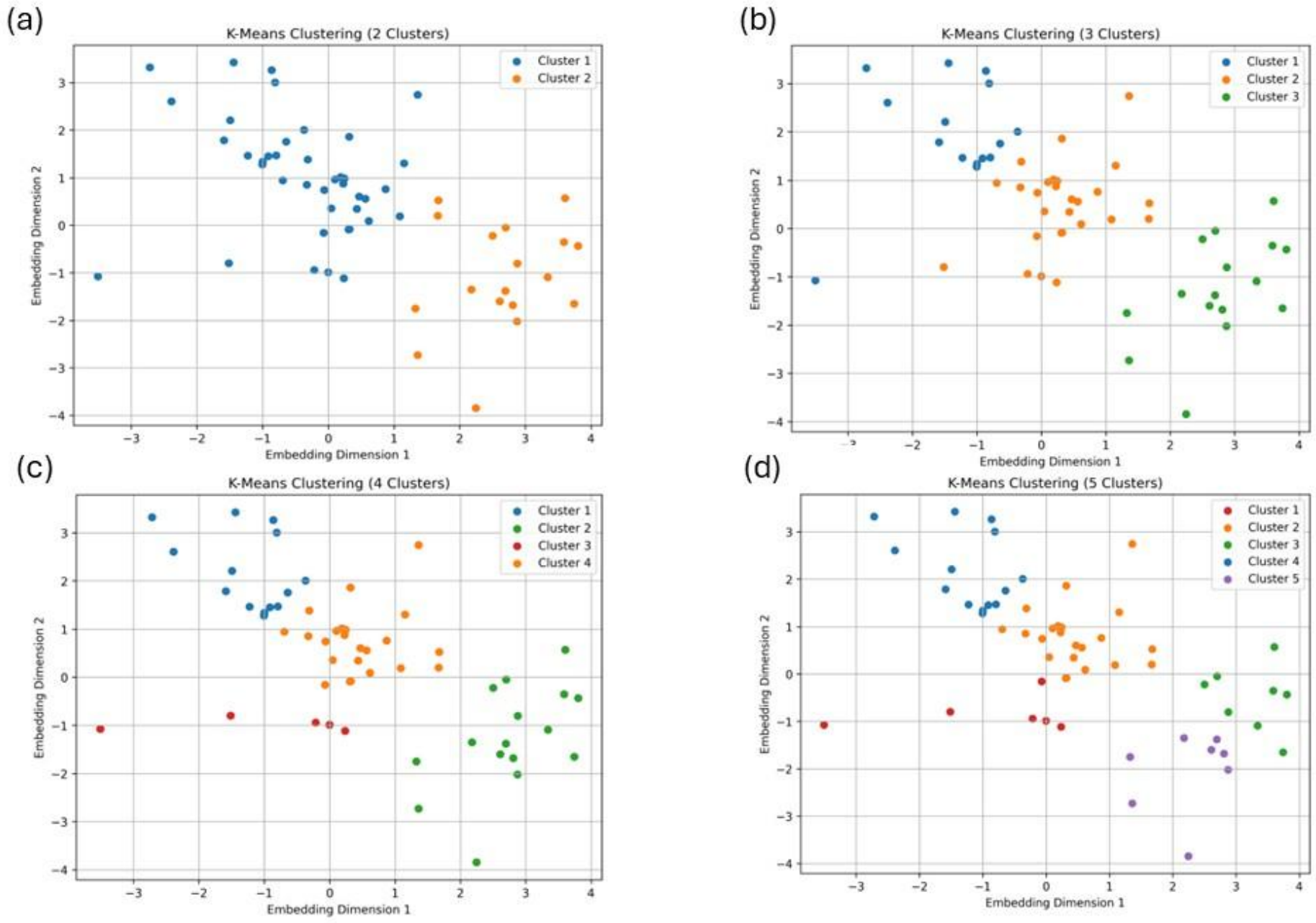

**Figure 4.** Two-dimensional representation of the learned embeddings for the 58 studies included in the dataset colored according to the cluster assignments obtained using $k$-means clustering for different predefined number of clusters (a)$k = 2$, (b) $k = 3$, (c) $k = 4$, and (d) $k = 5$.

## 4 Discussion

In this work, we proposed a novel framework that combines large language models with deep metric learning to explore heterogeneity in meta-analyses of observational studies. An LLM was used to extract and encode rich study-level information in the form of study triplets, which were subsequently incorporated into an embedding model trained with triplet loss. This approach enables the construction of an embedding space that captured similarities between studies and facilitated the identification of related study groups prior to statistical synthesis.

We applied the proposed framework to a meta-analysis of 58 observational studies measuring the IQ differences between children born pre-term and those born at full term. Traditional approaches including subgroup analyses and meta-regressions, were unable to adequately explain the observed between-study heterogeneity, thereby limiting the interpretability and reliability of pooled estimates. In contrast, the proposed framework generated an embedding space that captured clinically and methodologically meaningful similarities between studies, with subsequent $k$-means clustering ($k = 3$) identifying three principal study subgroups. Although Clusters 2 and 3 yielded pooled estimates similar to each other and to the overall analysis and retained considerable residual heterogeneity, Cluster 1 was characterized by a more extreme pooled SMD, lower between-study heterogeneity, and narrower prediction intervals. Sensitivity analyses based on alternative clustering ($k = 4$ and $k = 5$) consistently reproduced a cluster highly similar to Cluster 1 of the primary analysis, supporting the stability of this subgroup. Additional sensitivity analyses further demonstrated robustness of the findings to alternative triplet selections and tuning parameter specifications.

Taken together, these findings suggest that the proposed methodology identifies a latent representation that captures clinically and methodologically meaningful structures not apparent from meta-analytic summaries alone. In particular, Cluster 1 appeared to represent a subgroup of the most homogeneous studies reporting more consistently severe cognitive deficits among children born pre-term, whereas Clusters 2 and 3 reflected broader and more heterogeneous study populations. Importantly, Cluster 1 included 15 studies, a sample size sufficient for robust inference in most evidence synthesis frameworks.[11]. However, the extent to which these more pronounced pooled effect sizes translate to clinically important differences remains to be established. Consequently, further investigation into the underlying drivers within Cluster 1 may provide critical insights to refine real-world clinical practices.

A key advantage of the proposed framework is its ability to incorporate study-level information at an early stage of evidence synthesis process. By integrating pre-specified covariates into the LLM-driven construction of training triplets, the framework preserves domain knowledge while enabling the identification of groups of similar studies prior to statistical synthesis. Unlike traditional meta-regression, which incorporates covariates after synthesis to explain between-study heterogeneity, the proposed methodology uses study characteristics to organize evidence a priori. Hence, the developed approach may be combined with meta-regression. After identifying more homogeneous study subgroups, conventional meta-regression analyses may subsequently be performed within individual clusters to further investigate residual heterogeneity. Alternative meta-analysis models can also induce clusters, but these are primarily driven by similarities in effect sizes[33–35]. As a result, determining whether such clusters correspond to meaningful clinical or methodological differences between studies remains a separate, post hoc analysis. In contrast, our framework shifts the role of covariates from retrospective explanation of heterogeneity to pre-synthesis structuring of the evidence base, providing a complementary perspective for further investigating sources of variability among studies.

This study comes with some limitations. First, the framework depends on the quality and consistency of information extracted by the LLM, which may be sensitive to the prompt design and model specification. Future research should explore alternative prompting strategies[36] and the use of multiple LLMs to assess the robustness of the extracted information. Second, given the novelty of integrating LLMs with metric learning in meta-analysis, no gold-standard methodology yet exists for evaluating LLM-generated triplets. In the present work, evaluation relied primarily on manual review of a random subset of LLM-generated explanations for the constructed triplets. Therefore, the development of formal validation procedures is needed, including statistical stopping criteria for automated screening[37], to determine a sufficient subset of explanations required for reliable triplet evaluation. Third, while we used a relatively large number of studies, meta-analyses often involve significantly smaller datasets, which may increase the risk of overfitting in deep metric learning models. Although clustering with $k$-meas consistently identified a subgroup comprising the most homogeneous studies, future research should also investigate alternative clustering algorithms to assess the robustness of the identified study groupings[38–40]. In addition, systematic comparisons between the proposed framework and more flexible meta-analysis models designed to capture latent clustering structures, such as Dirichlet process approaches[33–35], would provide a valuable insight into the relative strengths and limitations of different methodologies for investigating heterogeneity in evidence synthesis.

In conclusion, this study demonstrates that integrating LLM-derived study representations with deep metric learning provides a promising and flexible framework for exploring heterogeneity in meta-analysis. The proposed approach may be particularly useful for synthesizing non-randomized evidence, where complex clinical and methodological variability is common and often insufficiently addressed by conventional methods. By leveraging study-level characteristics, the framework offers a data-driven mechanism for identifying more homogeneous subgroups of studies prior to quantitative synthesis. Conducting separate meta-analyses within these subgroups has the potential to reduce between-study heterogeneity and improve the interpretability of pooled estimates, thereby offering a more reliable evidence base for informed medical decision-making.

**Acknowledgments**

Theodoros Evrenoglou was supported by the Deutsche Forschungsgemeinschaft (DFG, German Research Foundation) under the Project ID-554095932.

The work of Harald Binder was funded by the Deutsche Forschungsgemeinschaft (DFG, German Research Foundation) – Project-ID 499552394 – SFB 1597